\input macros.sty

\def\ra{\rightarrow}

\begin{ifnosubmit}
\end{ifnosubmit}

\documentstyle[prd,aps,floats,epsfig,verbatim,mymcite,amsbsy]{revtex}

\graphicspath{{figures/}}

\newdimen\figsize
\newbox\figbox

\preprint{Fermilab-Conf-99/206-E}
\begin{document}

\preprint{Fermilab-Conf-99/206-E}

\begingroup
\large

\title{\Large \bf Neural Networks for Analysis of Top Quark Production}
\author{The \dzero\ Collaboration
\footnote{Submitted to the \textit{International Europhysics Conference
on High Energy Physics, EPS-HEP99}, 15--21 July, 1999, Tampere, Finland.}}
\address{Fermi National Accelerator Laboratory, Batavia, Illinois, 60510}
\date{July 15, 1999}
\maketitle

\vskip 1cm
\centerline{Abstract}
\vskip 1cm

\begin{abstract}
Neural networks (NNs) provide a powerful and flexible tool
for selecting a signal from a larger background.  The \dzero\ collaboration
has used them extensively in studying $t\bar t$ decays.
NNs were essential to the measurement of the $\ttbar$ production
cross section in the all-jets
channel ($t\bar t\rightarrow b\bar b qqqq$), and were also used in
the measurement of the mass of the top~quark
in the lepton+jets channel
($t\bar t\rightarrow b\bar b\ell\nu q\bar q$).
This paper will describe two new applications of neural networks to
top~quark analysis: the search for single top~quark production, and an
effort to increase the sensitivity in the dilepton channel
$t\bar t\rightarrow b\bar b e\bar\mu\nu\bar\nu$
beyond that achieved in the published analysis.
\end{abstract}

\endgroup

\newpage
{\def\author#1{#1}\def\address#1{#1}\raggedright
\author{                                                                      
B.~Abbott,$^{45}$                                                             
M.~Abolins,$^{42}$                                                            
V.~Abramov,$^{18}$                                                            
B.S.~Acharya,$^{11}$                                                          
I.~Adam,$^{44}$                                                               
D.L.~Adams,$^{54}$                                                            
M.~Adams,$^{28}$                                                              
S.~Ahn,$^{27}$                                                                
V.~Akimov,$^{16}$                                                             
G.A.~Alves,$^{2}$
% ...
\\                                                                            
\vskip 0.30cm                                                                 
\centerline{(D\O\ Collaboration)}                                             
\vskip 0.30cm                                                                 
}                                                                             
\address{                                                                     
\centerline{$^{1}$Universidad de Buenos Aires, Buenos Aires, Argentina}       
\centerline{$^{2}$LAFEX, Centro Brasileiro de Pesquisas F{\'\i}sicas,         
                  Rio de Janeiro, Brazil}                                     
\centerline{$^{3}$Universidade do Estado do Rio de Janeiro,                   
                  Rio de Janeiro, Brazil}                                     
\centerline{$^{4}$Institute of High Energy Physics, Beijing,                  
                  People's Republic of China}                                 
\centerline{$^{5}$Universidad de los Andes, Bogot\'{a}, Colombia}             
\centerline{$^{6}$Universidad San Francisco de Quito, Quito, Ecuador}         
\centerline{$^{7}$Institut des Sciences Nucl\'eaires, IN2P3-CNRS,             
                  Universite de Grenoble 1, Grenoble, France}                 
\centerline{$^{8}$DAPNIA/Service de Physique des Particules, CEA, Saclay,     
                  France}                                                     
\centerline{$^{9}$Panjab University, Chandigarh, India}                       
\centerline{$^{10}$Delhi University, Delhi, India}                            
\centerline{$^{11}$Tata Institute of Fundamental Research, Mumbai, India}     
\centerline{$^{12}$Kyungsung University, Pusan, Korea}                        
\centerline{$^{13}$Seoul National University, Seoul, Korea}                   
\centerline{$^{14}$CINVESTAV, Mexico City, Mexico}                            
\centerline{$^{15}$Institute of Nuclear Physics, Krak\'ow, Poland}            
\centerline{$^{16}$Institute for Theoretical and Experimental Physics,        
                   Moscow, Russia}                                            
\centerline{$^{17}$Moscow State University, Moscow, Russia}                   
\centerline{$^{18}$Institute for High Energy Physics, Protvino, Russia}       
\centerline{$^{19}$Lancaster University, Lancaster, United Kingdom}           
\centerline{$^{20}$University of Arizona, Tucson, Arizona 85721}              
\centerline{$^{21}$Lawrence Berkeley National Laboratory and University of    
                   California, Berkeley, California 94720}                    
\centerline{$^{22}$University of California, Davis, California 95616}         
\centerline{$^{23}$University of California, Irvine, California 92697}        
\centerline{$^{24}$University of California, Riverside, California 92521}     
\centerline{$^{25}$Florida State University, Tallahassee, Florida 32306}      
\centerline{$^{26}$University of Hawaii, Honolulu, Hawaii 96822}              
\centerline{$^{27}$Fermi National Accelerator Laboratory, Batavia,            
                   Illinois 60510}                                            
\centerline{$^{28}$University of Illinois at Chicago, Chicago,                
                   Illinois 60607}                                            
\centerline{$^{29}$Northern Illinois University, DeKalb, Illinois 60115}      
\centerline{$^{30}$Northwestern University, Evanston, Illinois 60208}         
\centerline{$^{31}$Indiana University, Bloomington, Indiana 47405}            
\centerline{$^{32}$University of Notre Dame, Notre Dame, Indiana 46556}       
\centerline{$^{33}$Purdue University, West Lafayette, Indiana 47907}          
\centerline{$^{34}$Iowa State University, Ames, Iowa 50011}                   
\centerline{$^{35}$University of Kansas, Lawrence, Kansas 66045}              
\centerline{$^{36}$Kansas State University, Manhattan, Kansas 66506}          
\centerline{$^{37}$Louisiana Tech University, Ruston, Louisiana 71272}        
\centerline{$^{38}$University of Maryland, College Park, Maryland 20742}      
\centerline{$^{39}$Boston University, Boston, Massachusetts 02215}            
\centerline{$^{40}$Northeastern University, Boston, Massachusetts 02115}      
\centerline{$^{41}$University of Michigan, Ann Arbor, Michigan 48109}         
\centerline{$^{42}$Michigan State University, East Lansing, Michigan 48824}   
\centerline{$^{43}$University of Nebraska, Lincoln, Nebraska 68588}           
\centerline{$^{44}$Columbia University, New York, New York 10027}             
\centerline{$^{45}$New York University, New York, New York 10003}             
\centerline{$^{46}$University of Rochester, Rochester, New York 14627}        
\centerline{$^{47}$State University of New York, Stony Brook,                 
                   New York 11794}                                            
\centerline{$^{48}$Brookhaven National Laboratory, Upton, New York 11973}     
\centerline{$^{49}$Langston University, Langston, Oklahoma 73050}             
\centerline{$^{50}$University of Oklahoma, Norman, Oklahoma 73019}            
\centerline{$^{51}$Brown University, Providence, Rhode Island 02912}          
\centerline{$^{52}$University of Texas, Arlington, Texas 76019}               
\centerline{$^{53}$Texas A\&M University, College Station, Texas 77843}       
\centerline{$^{54}$Rice University, Houston, Texas 77005}                     
}                                                                             
}

\twocolumn\narrowtext


\section{Introduction}

Since the observation
of the top~quark in~1995\mcite{cdfdiscovery,*d0discovery},
much experimental effort has been invested
in studying its properties\cite{review}.  Such analyses are difficult, owing
to the small number of $\ttbar$ events available, the relatively
large backgrounds, and the complex event geometries.  There
has therefore been a great deal of interest in analysis techniques that could
improve on the standard methods of selecting candidate events.  One useful
class of such techniques uses pattern classifiers based on
feed-forward ``neural networks.'' \mcite{rumelhart86,*beale90}

The \dzero\ experiment at the Fermilab Tevatron has made considerable
use of neural network techniques in its analyses of top quark data.  Both the
cross section measurement in the all-jets
channel\mcite{d0alljetsprd,*d0alljetsprl} and the mass
measurement in the lepton + jets channel\mcite{d0ljtopmassprd,*d0ljtopmassprl}
used neural networks;
details of these analyses have already been published.

Here, we describe two more recent studies: a neural network
analysis of single top quark production, and an effort to improve the
efficiency for selecting $\ttbar \ra e\mu$ events using neural
networks.  We shall start with a brief description of the kind of neural
networks used in these analyses.

\section{Neural Networks}

\begin{figure}
\centering
\figsize=\hsize
\advance\figsize by -3cm
\epsfig{file=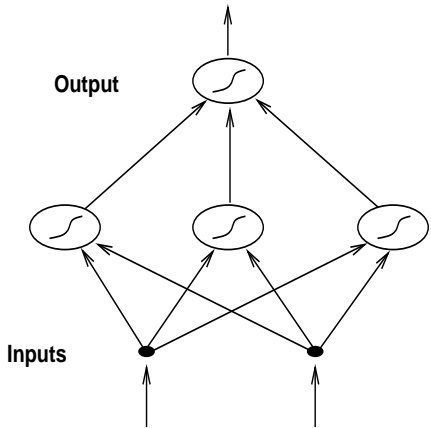, width=\figsize, height=\figsize}
\caption{A feed-forward neural network.}
\label{fg:nndiag}
\end{figure}

\Figref{fg:nndiag} shows an example of the type of neural network used 
in these studies.  It consists of a set of processing units, each
of which has at least one input and one output.  The output $y_i$ of a
single unit $i$ is given in terms of its inputs $x_{ij}$ by
\begin{equation}
  y_i = g (\sum_j x_{ij} + \theta_i),
\end{equation}
where $\theta_i$ is a threshold specific to the unit,
and $g$ is a nonlinear squashing 
function, typically of the form
\begin{equation}
  g(x) = {1 \over 1 + e^{-2x}}.
\end{equation}
[Thus, the unit outputs are bounded in the range $(0,1)$.]
The units are arranged in layers, with the inputs of layer $n+1$
connected to the outputs of layer $n$ by a weight matrix:
\begin{equation}
  x^{n+1}_{ij} = w_{ij} y^n_j.
\end{equation}
Typically, the last layer consists of only one unit, and is called the
``output'' layer; the other layers are called ``hidden'' layers.
Often, the $y_i^0$ are said to be the outputs of a dummy ``input''
layer.  No processing, however, is done in that ``layer.''
Such a network is quite flexible; in fact, it has been shown that a
network with only one hidden layer can approximate any reasonable
(Borel-measurable) function to any required degree of accuracy,
provided that sufficient units are available in the
hidden layer\mcite{hornik89,*funahashi89}.

For pattern recognition, one wants to have the network
output~1 if the input is most consistent with signal, and~0 if the
input is most consistent with background.  Typically, one has
available a collection of $N$ inputs, some of which are known to be signal 
and some of which are known to be background.  One defines an error
function:
\begin{equation}
  \chisq = {1\over N}\sum_{i=1}^N (O_i - t_i)^2,
\end{equation}
where $O_i$ is the output of the network for input $i$, and $t_i$ is
the desired output for that input.  This quantity can be considered as 
a function of the weights ${\bf w}$ and thresholds ${\boldsymbol\theta}$; one
then minimizes $\chisq$ with respect to these variables to achieve an
approximation to the desired function.

The minimization technique most often used is called
``backpropagation,'' which is a sort of stochastic gradient
descent.  Other minimization algorithms can also be used.
This process is often referred to as ``training''
the network.

\section{Single Top Quark Production}

\begin{figure}
\centering
\epsfig{file=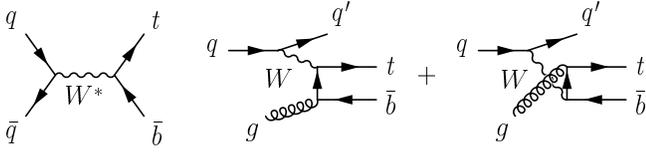, width=\hsize}
\caption{Feynman diagrams for single top~quark production.}
\label{fg:singletop-prod}
\end{figure}

The first study we will examine is a search for single top quark
production \cite{lev}.  The processes relevant at the Tevatron are
illustrated in \figref{fg:singletop-prod}; the total cross sections
for these processes calculated at next-to-leading order (NLO)
are \mcite{smith96,*stelzer97,*heinson97b}:
\begin{eqnarray}
  \sigma_{\text{NLO}} (\ppbar \ra t\bbar X + \text{c.c.}) &=
      0.724 \pm 0.043\pb,\\
  \sigma_{\text{NLO}} (\ppbar \ra tq\bbar X + \text{c.c.}) &=
      1.70 \pm 0.27\pb.\nonumber
\end{eqnarray}
Such processes are interesting because they directly probe the
$W-t-b$ vertex.  Assuming the Standard Model, measuring these cross
sections gives a measurement of the $V_{tb}$ element of the
Cabibbo-Kobayashi-Maskawa (CKM) matrix.  Such measurements are also
sensitive to any new physics in the weak interactions of the
top~quark\cite{tait97}.

After the decay of the top~quark, the particles produced in these
processes are $Wb\bbar$ and $Wb\bbar q$, possibly with additional jets 
from QCD radiative effects.  This study looks for leptonic decays of
the $W$~boson, so the initial event selection requires a high-$\pt$
lepton, large missing transverse energy ($\met$), and at least two
jets.  No $b$-tag is required in this study, in order to preserve
signal efficiency (but if information about a tagging muon is
present, it will be used).

The numbers of signal and background events expected to remain after this
selection for \dzero's Run~1 ($109\ipb$) are as follows:

\medskip
{\centering
\vbox{\begin{tabular}{|ll|}
\hline
  Process & $N_{\text{events}}$\\
\hline
\hline
  $tb$  & $2.1$ \\
  $tqb$ & $5.1$ \\
\hline
\hline
  QCD multijet\quad\quad & 2411 \\
  $\ttbar$     & 22.3 \\
  $Wbb$ & 11.4 \\
  $Wjj$ ($c$, $s$) & 51.8 \\
  $Wjj$ ($g$, $u$, $d$) & 1615.7 \\
  $WW$                     & 36.9 \\
  $WZ$                     & 5.3\\
\hline
\end{tabular}}}
\medskip

As can be seen, the background is huge compared to the signal, with
the dominant background sources being QCD multijet production (with a
jet misidentified as a lepton) and the production of $W$~bosons with
associated jets.

A crucial step in a neural network analysis is the selection of
the variables used as input to the network.
Adding more variables potentially 
increases the amount of information available to the network, but it
also expands the space that must be searched during the minimization, 
making it more difficult to find a good minimum.  In fact, with some
procedures, adding variables of marginal utility can degrade
the performance of a network.  And while neural networks can in
principle approximate any reasonable function, in practice
complicated mappings may require too many hidden nodes
for minimization to be practical.

\begin{figure}
\centering
\vbox{%
\epsfig{file=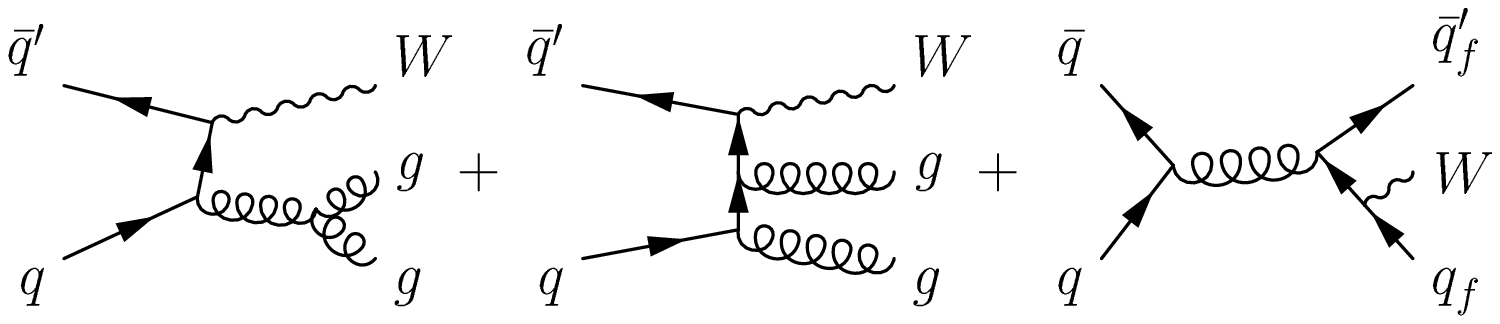, width=\hsize}
\vskip 0.5cm
\epsfig{file=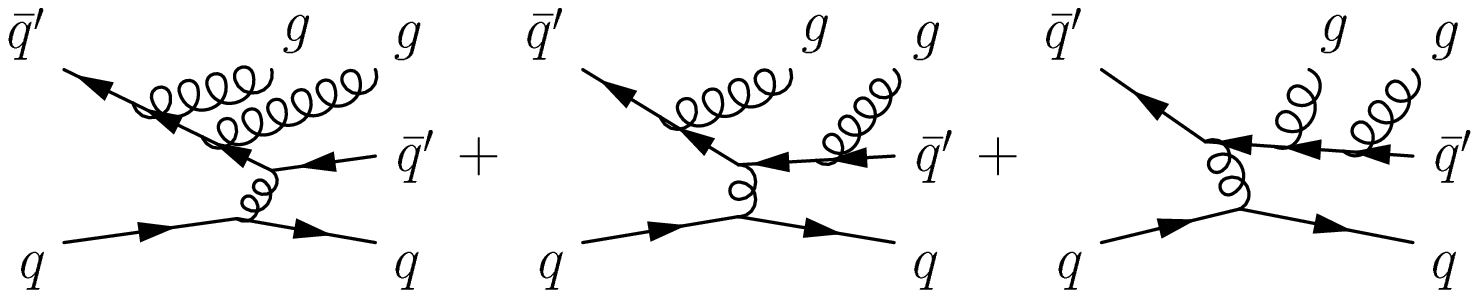, width=\hsize}}
\vskip 0.3cm
\caption{Typical Feynman diagrams for the $Wjj$ and QCD backgrounds
  to single top~quark production.}
\label{fg:singletop-bg}
\end{figure}

A useful observation is that the rate for a scattering process is
greatest in the regions of phase space near singularities in the
corresponding matrix element\mcite{boos99}.
If such singularities occur in
different places for signal and background, then the
dependence on the corresponding variables 
in which the singularities occur should differ strongly between signal
and background.  For example, the top~quark production diagrams
in \figref{fg:singletop-prod} have a singularity at
$M_t^2 = (p_b + p_W)^2 \ra \mt^2$.  In contrast, the dominant background
diagrams, illustrated in \figref{fg:singletop-bg}, have singularities at
\begin{eqnarray}
M_{g1, g2}^2 &= (p_{g1} + p_{g2})^2 \ra 0, \\
\hat t_{q,(g1g2)} &= (p_{g1} + p_{g2} - p_q)^2 \ra 0, \\
\hat t_{q,g1} &= (p_{g1} - p_q)^2 \ra 0, \\
\hat t_{q,g2} &= (p_{g2} - p_q)^2 \ra 0.
\end{eqnarray}
These variables, however, are defined at the parton level, and cannot
be directly measured,
due to effects of QCD radiation, the
unobserved neutrino, and unobserved momentum that escapes down
the beam pipe.  In such a situation, it is better to use other
variables that are related to the singular variables, but can be
derived directly from the observed final state.  For example,
the typical $t$-channel singular variable $\hat t_{i,f}$ associated
with the production of a light particle (or jet) $f$ can be written
\begin{equation}
  \hat t_{i,f} = (p_f - p_i)^2 = - \sqrt{\hat s} e^{Y} \pt^f e^{-|y_f|},
\end{equation}
where $\sqrt{\hat s}$ is the total invariant mass of the produced system, $Y$ 
is its total rapidity, and $\pt^f$ and $y_f$ are the transverse
momentum and rapidity of the produced $f$.

\begin{figure}
\centering
\epsfig{file=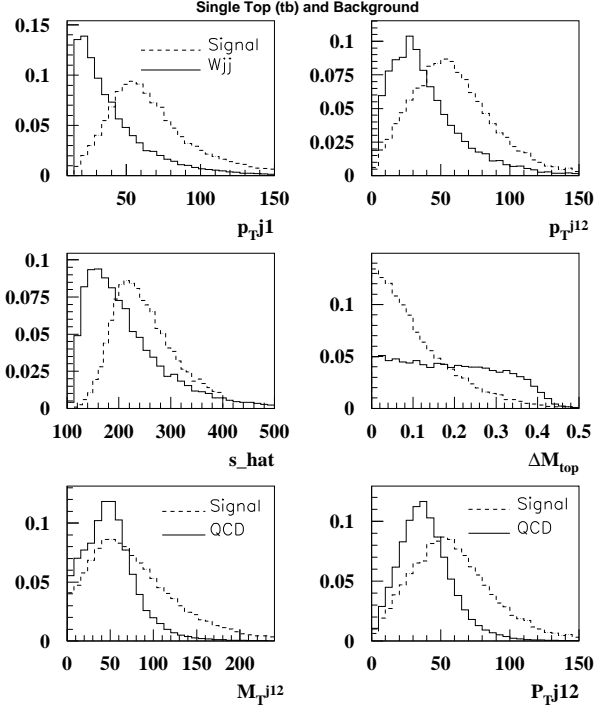, width=\hsize}
\caption{Distributions of kinematic variables for single~top signal
  (dashed) and background (solid), either $Wjj$ (top four) or QCD
  (bottom two).  Units are in $\ugevcc$.}
\label{fg:singletop-vars}
\end{figure}

From these kinds of considerations, a nominal set of input variables
can be defined as:
\begin{eqnarray}
  \text{Set 1: }&\text{$M_{j1, j2}$, $M_t$, $Y_{\text{tot}}$, ${\pt}_{j1}$, $y_{j1}$,}\\
  &\text{${\pt}_{j2}$, $y_{j2}$, ${\pt}_{j12}$, $y_{j12}$, $\sqrt{\hat s}$},\nonumber
\end{eqnarray}
where ${\pt}_{j12}$ and $y_{j12}$ are the transverse momentum and
rapidity of the system formed by the two highest $\pt$ jets, and
$Y_{\text{tot}}$ is the total rapidity of the center of mass of
the initial partons, as reconstructed from the final state.  The
$z$-component of the momentum of the $W$~boson is found by enforcing the
$M_W$ mass constraint in the leptonic $W$~boson decay.  
Distributions of some of these
variables are shown in \figref{fg:singletop-vars}.

\begin{figure}
\centering
\epsfig{file=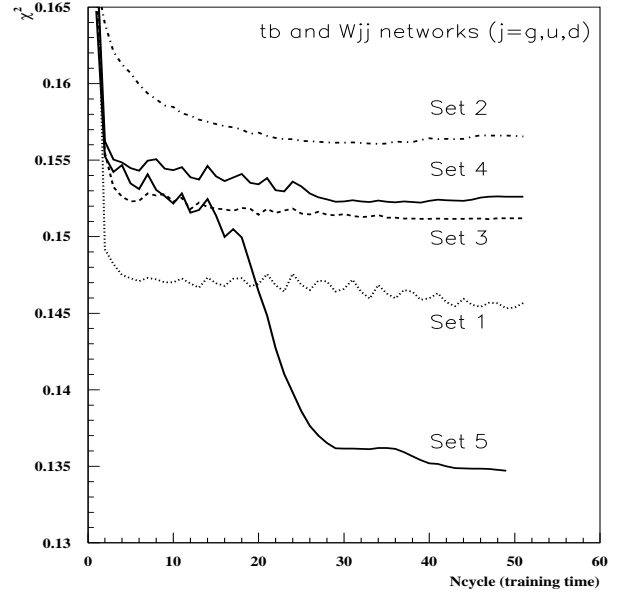, width=\hsize, height=\hsize}
\caption{Neural network $\chisq$ vs. training time, for different sets 
  of input variables.  The networks are trained on a sample consisting 
  of the single~top signal and $Wjj$ background.}
\label{fg:singletop-training}
\end{figure}

\Figref{fg:singletop-training} compares this set with the simpler
sets:
\begin{eqnarray}
  \text{Set 2: }&\text{${\pt}_{j1}$, ${\pt}_{j2}$,
                $H_{\text{all}}$, $H_{T\text{all}}$;}\\
  \text{Set 3: }&\text{${\pt}_{j1}$, ${\pt}_{j2}$,
                $H_{\text{all}}$, $H_{T\text{all}}$, $M_t$;}
\end{eqnarray}
where $H_{\text{all}} = \sum E_f$ and
$H_{T\text{all}} = \sum {\et}_f$.  The comparison is made by training 
a neural network for each of the sets on a sample of events consisting
of top~quark signal plus $Wjj$ background.  It is seen that the neural
network built using Set~1 performs better than those using Set~2 or
Set~3.

\Figref{fg:singletop-training} also shows two other variations of the
set of input variables.  Set~4 is the
same as Set~1, except that the variables $H_{\text{all}}$ and
$H_{T\text{all}}$ are added.  It is seen that this does worse than
Set~1 --- the additional variables do not add enough
information to counteract the increase in the size of the minimization 
space.  Set~5 adds to Set~1 the widths $w_{\text{jet}}$
of the two jets and the $\pt$
of a $b$-tagging muon (set to zero if there is no such tag).  In this
case, the added variables help: Set~5 has a lower $\chisq$ than any of 
the others.

\begin{figure}
\centering
\epsfig{file=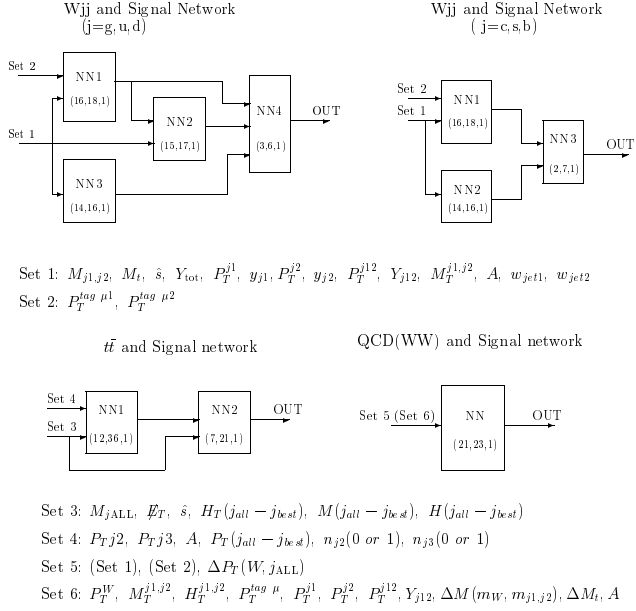, width=\hsize}
\caption{The structure of the neural networks used to reject each
  background.  Each network has one hidden layer; the notation
  $(n_1, n_2, n_3)$ gives the number of units in the input, hidden,
  and output layers of each network, respectively.  The following
  variables are used in addition to those defined in the text.
  $n_{j2}$ is 1 if the event has exactly two jets and 0 otherwise;
  $n_{j3}$ is 1 if the event has three or more jets and 0 otherwise.
  The jet for which the invariant mass of the lepton, neutrino, and
  jet is closest to $172\gevcc$ is denoted $j_{best}$; the notation
  $j_{all} - j_{best}$ means all jets except $j_{best}$. 
  Also, $\Delta\pt(W,j_{all}) = \pt(W) -
  \sum_{\text{jet}}\vec\pt^{\text{jet}}$,
  $\Delta M(m_W, m_{j1}, m_{j2}) = |m_W - m(j_1, j_2)| / m_W$, and
  $\Delta M_t = |m(e,\nu,j_{best}) - 172| / 172$.}
\label{fg:singletop-networks}
\end{figure}

\begin{figure}
\centering
\epsfig{file=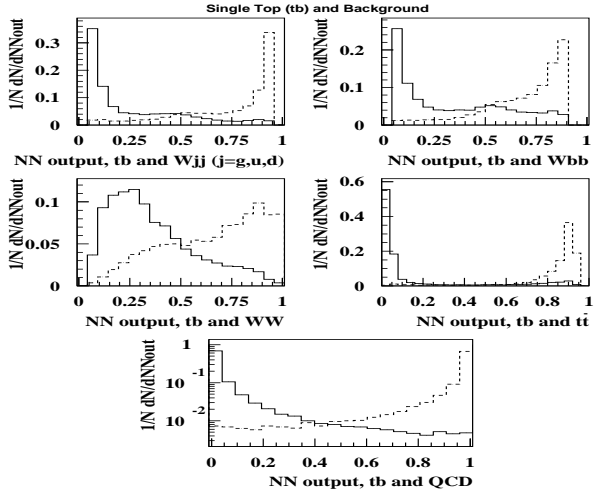, width=\hsize, height=0.8\hsize}
\caption{Outputs of each of the neural networks for single~top signal
  (dashed) and the indicated background (solid).}
\label{fg:singletop-netoutput}
\end{figure}

\begin{figure}
\centering
\epsfig{file=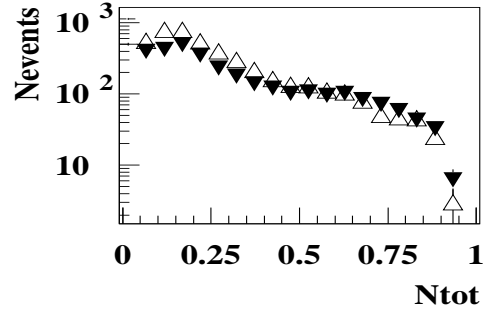, width=0.75\hsize, height=0.5\hsize}
\caption{A comparison of the combined output of the five networks for
  data (the open symbols) and a Monte Carlo model of signal and all
  backgrounds (the solid symbols).  The individual network outputs are
  combined using 
  ${1/O_{\text{tot}}} = (1/5)\sum_{i=1}^5 1/O_{\text{NN}i}$.}
\label{fg:singletop-datacomp}
\end{figure}

\begin{figure}
\centering
\epsfig{file=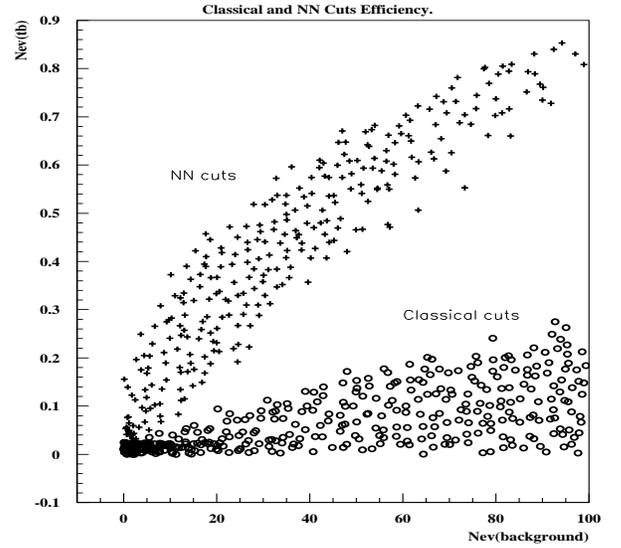, width=\hsize, height=0.9\hsize}
\caption{Comparison of signal/background efficiencies for NN and
  conventional analyses.  Each point represents one specific set of
  cuts.}
\label{fg:singletop-comparison}
\end{figure}

For the final analysis, a separate network is constructed for each of 
the major backgrounds, as shown in \figref{fg:singletop-networks}.
The networks are trained using \progname{jetnet}\cite{jetnet}; the
results for each network are shown in \figref{fg:singletop-netoutput}.
\Figref{fg:singletop-datacomp} shows that the network output
from Monte Carlo models agrees well with the data.  Finally, individual
cuts are made on each of the five network outputs.
\Figref{fg:singletop-comparison} compares the results of this
to a more conventional analysis.
It is seen that for a given 
background level, the neural network analysis provides several times
the signal efficiency of conventional cuts.

\section{$\ttbar$ Decays Into {$\lowercase{e}\mu$}}

The ``golden'' channel for observing $\ttbar$ decays has long been
the dilepton mode
$\ttbar \ra W^+bW^-\bbar \ra e \nu_e \mu \nu_\mu b\bbar$.  Due to the
presence of two leptons with different flavors, this channel has a very low
background.  However, compared to the channels in which one of the
$W$~bosons decays into jets, the $e\mu$ channel has a relatively
small branching ratio --- about $2.5\%$, versus about $15\%$ for the
$e+\jets$ channel.  Therefore, any new analysis techniques that can
increase efficiency for identifying signal in this
channel while maintaining the low background level are welcome.

This study starts from the published measurement of the $\ttbar$
production cross section\cite{d0xsecprl97}, which selects $e\mu$
candidates as follows:
\begin{itemize}
\item An electron with $\et > 15\gev$ and $|\eta| < 2.5$.
\item A muon with $\pt > 15\gevc$ and $|\eta| < 1.2$.
\item $\met > 20\gev$.
\item At least two jets with $\et > 20\gev$ and $|\eta| < 2.5$.
\item $\Delta R_{\mu,\jet} > 0.5$ and $\Delta R_{e,\mu} > 0.25$.
  ($\Delta R = \sqrt{(\Delta\phi)^2 + (\Delta\eta)^2}$.)
\item $H_T > 120\gev$, where $H_T = \et^e + \sum_{\jets} \et^{\jet}$.
\end{itemize}

For the present study, this selection is relaxed by removing
the cut on $H_T$ and reducing the $\met$ and jet $\et$ cuts to
$15\gev$.  This defines the sample used as input to the neural
network.

\begin{figure}
\centering
\epsfig{file=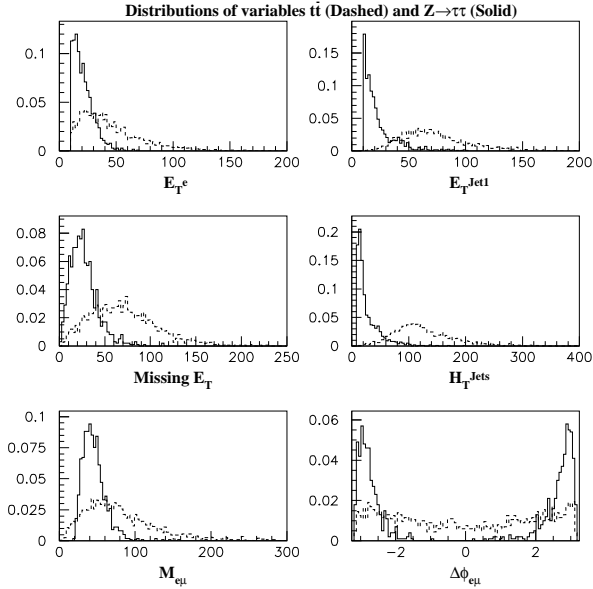, width=\hsize}
\caption{Distributions of input variables to the $\tau\tau$ neural
  network for signal (dashed) and $Z\ra\tau\tau$ background (solid).}
\label{fg:emu-vars}
\end{figure}

\begin{figure}
\centering
\epsfig{file=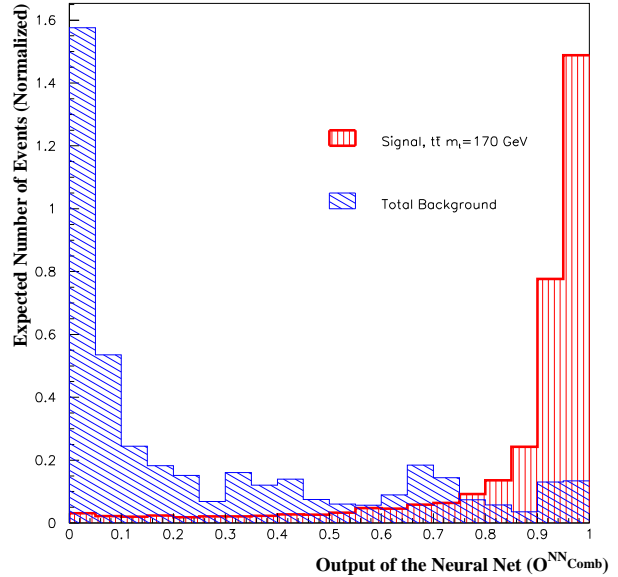, width=\hsize}
\caption{Distribution of $O_{\text{NN}}^{\text{comb}}$ for $\ttbar$
  signal and background events.}
\label{fg:emu-nncomb-sigbkg}
\end{figure}

There are three major backgrounds to contend with: QCD jet production
with jets misidentified as leptons, $Z\ra \tau\tau \ra e\mu$, and
$WW\ra e\mu$ events.  A separate network is trained to separate the signal
from each of the three backgrounds.  Six variables are used as inputs
to each of the networks, these being $\et^e$, $\et^{\jet 2}$, $\met$,
$H_T^{\jets} = \sum_{\jets} \et^{\jet}$, $M_{e\mu}$, and
$\Delta\phi_{e\mu}$, except for the $\tau\tau$ network, where
$\et^{\jet1}$ replaces $\et^{\jet2}$.  The input variables for the
$\tau\tau$ network are plotted in \figref{fg:emu-vars}.  
Each network has seven hidden units.  The networks
are trained (using \progname{jetnet})
on equal numbers of $\ttbar$ signal and background events 
(2000 of each for the QCD network, and 1000 of each for the other
two).  The outputs of the three networks are combined, as usual, using
\begin{equation}
  O_{\text{NN}}^{\text{comb}} = {3 \over { 1 \over O_{\text{NN}1} } +
                                         { 1 \over O_{\text{NN}2} } +
                                         { 1 \over O_{\text{NN}3} } }.
\end{equation}
Distributions of this variable for signal and background are shown in
\figref{fg:emu-nncomb-sigbkg}.  To define
the candidate sample, a final cut of
$O_{\text{NN}}^{\text{comb}} > 0.88$ is imposed, which was determined
by maximizing the expected relative significance, $S/\sigma_B$.
($\sigma_B$ is the uncertainty in the background estimate.)

\begin{figure}
\centering
\epsfig{file=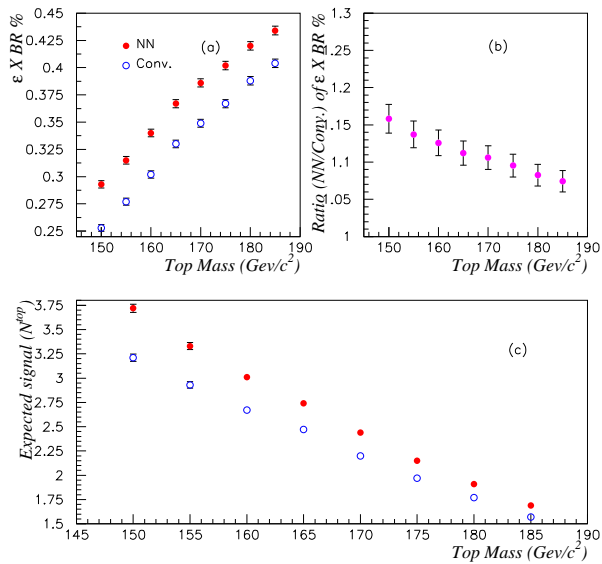, width=\hsize}
\caption{The neural network analysis compared to the standard
  analysis, for \dzero's Run~1.  (a) Efficiency times branching ratio
  (\%); (b) Ratio of NN analysis efficiency to standard analysis
  efficiency; (c) Expected number of signal events.
  Uncertainties displayed are statistical only; the systematic uncertainties
  (included in \tabref{tb:emu-results}) are highly correlated between
  the two analyses.}
\label{fg:emu-results}
\end{figure}

\begin{figure}
\centering
\epsfig{file=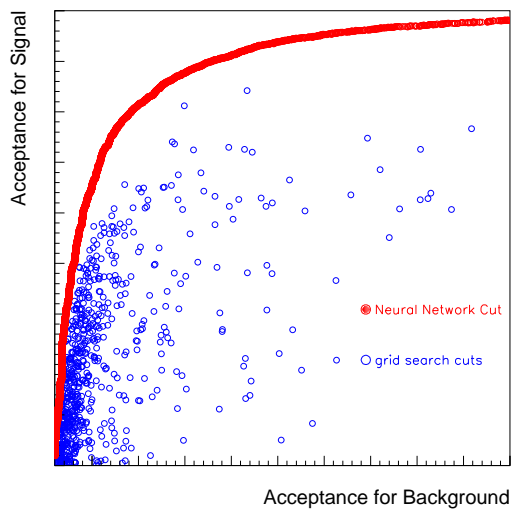, width=\hsize}
\caption{The neural network analysis compared to the standard
  analysis.  Each point represents a different set of selection
  requirements.}
\label{fg:emu-grid}
\end{figure}

The resulting signal efficiencies and estimated backgrounds for
\dzero's Run~1 ($108\ipb$) are shown in \tabref{tb:emu-results} and
\figref{fg:emu-results}.  Compared to the standard (published)
analysis, it is
seen that the neural network analysis increases the signal efficiency
by about $10\%$.  In addition, the background is also slightly lower,
although this is harder to evaluate due to the large statistical
errors in the QCD background sample.  Further comparison is made
in \figref{fg:emu-grid}.

\begin{table}
\centering
\begin{tabular}{ccc}
 & Conventional & NN\\
 & analysis     & analysis\\
\hline
\hline
Signal  & \multicolumn{2}{c}{$\epsilon\times\text{BR}$ (\%)}\\
\hline
$\mt=170\gevcc$ & $0.349\pm 0.074$ & $0.386\pm 0.082$ \\
$\mt=175\gevcc$ & $0.368\pm 0.078$ & $0.402\pm 0.085$ \\
$\mt=180\gevcc$ & $0.388\pm 0.082$ & $0.420\pm 0.089$ \\
\hline
\hline
Background & \multicolumn{2}{c}{$N_{\text{expected}}$}\\
\hline
$Z\ra \tau\tau \ra e\mu$ & $0.10\pm 0.10$ & $0.10\pm 0.07$ \\
$WW \ra e\mu$ & $0.074\pm 0.020$ & $0.085\pm 0.023$ \\
$\gamma^*\ra \tau\tau \ra e\mu$ & $0.006\pm 0.005$ & $0.007\pm 0.006$ \\
Fakes & $0.083\pm 0.126$ & $0.048\pm 0.124$ \\
Total & $0.26\pm 0.16$ & $0.24\pm 0.15$ \\
\end{tabular}
\caption{A comparison of the results of the conventional and neural
  network $\ttbar\ra e\mu$ analyses.  The numbers of background events
  are normalized for \dzero's Run~1 ($108\ipb$).}
\label{tb:emu-results}
\end{table}

\section{Conclusions}

In both the analyses considered here, neural networks provide a
significant improvement over conventional analysis methods.  We expect 
that such techniques will have a prominent place in the analysis of
data from the upcoming Run~2 of the Tevatron.

\section*{Acknowledgments}
%
We thank the Fermilab and collaborating institution staffs for
contributions to this work and acknowledge support from the 
Department of Energy and National Science Foundation (USA),  
Commissariat  \` a L'Energie Atomique (France), 
Ministry for Science and Technology and Ministry for Atomic 
   Energy (Russia),
CAPES and CNPq (Brazil),
Departments of Atomic Energy and Science and Education (India),
Colciencias (Colombia),
CONACyT (Mexico),
Ministry of Education and KOSEF (Korea),
and CONICET and UBACyT (Argentina).

\bibliographystyle{reviewsty}
\bibliography{nnnote}

\end{document}